\documentclass[prl,twocolumn,showpacs,letterpaper]{revtex4}
\usepackage[dvips]{graphics,graphicx}
\usepackage{amssymb}
\usepackage{bm}
\usepackage{amsmath}

\newcommand{\OH}{{\hat{H}}}

\newcommand{\OP}{{\hat{\ef}}}
\newcommand{\Oa}{{\hat{a}}}
\newcommand{\Oaa}{{\hat{a}^\dagger}}

\newcommand{\On}{{\hat{n}}}
\newcommand{\ef}{{\Phi}}
\newcommand{\mf}{m^\text{eff}}
\newcommand{\tm}{\bar \tau}
\newcommand{\Knn}{\zeta}
\newcommand{\Kab}{\kappa}
\newcommand{\lf}{\epsilon}
\newcommand{\oB}{\Delta}
\renewcommand{\l}{{\ell}}
\renewcommand{\t}{{\tau}}
\newcommand{\ket}[1]{|#1\rangle}
\newcommand{\ex}[1]{\langle#1\rangle}
\newcommand{\scal}[2]{\ensuremath{\langle #1 | #2 \rangle}}
\newcommand{\diff}[2]{\frac{\partial #1}{\partial #2}}
\newcommand{\Diff}[2]{\frac{{d} #1}{{d} #2}}

\newcommand{\N}{\mathbb{N}}
\renewcommand{\H}{\mathcal{H}}
\newcommand{\R}{\mathbb{R}}
\newcommand{\sgn}{{\mathrm{sgn}}}

\begin{document}
\title{Oscillation dynamics of multi-well condensates}
\author{S. Mossmann}
\email{mossmann@fis.unam.mx}
\author{C. Jung}
\affiliation{Instituto de Ciencias F\'{\i}sicas, Universidad Nacional Aut\'onoma de M\'exico,
 62251 Cuernavaca, M\'exico}
\date{\today }

\begin{abstract}
We propose a new approach to the macroscopic dynamics of three-well Bose-Einstein condensates, giving particular emphasis to self-trapping and Josephson oscillations.
Although these effects have been studied quite thoroughly in the mean-field approximation, a full quantum description is desirable, since it avoids pathologies due to the nonlinear character of the mean-field equations. Using superpositions of quantum eigenstates, we construct various oscillation and trapping scenarios.
\end{abstract}
\pacs{03.75.Kk, 03.75.Lm, 03.65.Sq}
\maketitle



Interference phenomena belong to the most surprising features of quantum
mechanics, which in many cases run against our classical everyday
experience. Most spectacular is this nonclassical behavior, if it
occurs in macroscopic systems. Such cases can help to sharpen our
eyes for quantum properties of matter. Therefore, it is always
worth to point out situations, in which new examples of macroscopic
nonclassical behavior might be accessible to laboratory experiments.
The best candidates for macroscopic quantum phenomena are superfluids
and condensates. This is certainly one motivation for the enormous
interest in condensates at the moment. In addition, condensates
serve as systems of analogy for many structures from very different
subbranches of physics. This makes one hope that phenomena
pointed out for condensates can find applications in various areas of physics.

In this letter, we present a new idea for a quantum oscillation of a
condensate in three wells. It is a system, for which we have already presented
a semiclassical analysis in \cite{06semiBEC}. The complete classification of the
eigenstates of the system in a dynamical representation makes it easy to select appropriate
eigenstates, which, when superimposed, result in a packet with the
desired dynamical properties.
We will show the following: By forming a packet from a few
(in this letter we only use two) appropriate
eigenstates, we produce oscillations between any pair of wells
we want. The most counterintuitive case
is an oscillation between the two outer wells, where the middle well
is almost empty and does not participate in the oscillations.

Macroscopic effects like self-trapping and Josephson oscillations have been 
investigated for the two-well potential \cite{Java86,Milb97,Smer97,Ragh99} and also for the three-well
potential \cite{Nemo00,Fran03}. However, the behavior has been studied mostly in the mean-field
approximation using the nonlinear Gross-Pitaevskii equation or for specific quantum states. The subsequent investigation is 
done thoroughly in the many particle quantum system avoiding all the difficulties related to the 
nonlinearity of the mean-field equations. 
Since the described phenomena do not require sophisticated preparation procedures,
they should be directly accessible to experimental investigations like, e.g., in 
Refs.~\cite{Albi05,Schu05a}.

We consider the situation, where a Bose-Einstein condensate is confined in a linear configuration of three potential wells. In experiments, this can be achieved, e.g., by combining an optical dipole trap with a standing laser wave \cite{Albi05}. 
We restrict the analysis in this letter to three wells. A convenient set of basis functions for such 
systems are the Wannier functions $\varphi_{k,l}(x)$ \cite{Kohn59}, which are real-valued and 
exponentially localized in each well. The index $k$ labels the potential sites, while $l$ denotes the 
excitation in each well. Since we focus on condensed particles, we can assume 
$l=0$ and skip the index in the following. Thus, the three Wannier functions 
$\{\varphi_{1},\varphi_2,\varphi_{3}\}$ span the one particle Hilbert space $\H_1$. For the 
subsequent discussion we only need the spacial localization properties of the basis, so that the 
analysis applies also to other physical situations like, e.g., atom chips \cite{Schu05a}. 

Expanding the field operator $\OP(x)$ in the one particle basis, $\OP(x) = \sum 
\varphi_k(x)\,\Oa_k$, yields in the standard tight binding approximation the many 
particle Hamiltonian 
\begin{align}
\OH =&\;\sum_{k=1}^3\lf_k\,\bigl(\Oaa_k\Oa_k+{\textstyle\frac{1}{2}}\bigr)
+\Knn\,\bigl(\Oaa_k\Oa_k+{\textstyle\frac{1}{2}}\bigl)^2 \nonumber\\
&\; -\Kab_{12}\,(\Oaa_1\Oa_2+\Oaa_2\Oa_1)
-\Kab_{23}\,(\Oaa_2\Oa_3+\Oaa_3\Oa_2)\,.\label{eq-Hqm}
\end{align}
The constants are given by
\begin{align}
\Kab_{kl} &= \int \varphi^*_k(x)\Bigl[-\frac{1}{2}\Diff{}{x} + V(x)\Bigr]\varphi_l(x)\,dx\,,\\ 
\lf_k &= \Kab_{kk} = \bar\lf +(k-2)\,\oB\,,\quad k=1,2,3\label{eq-WSL}\,,\\
\Knn &= {\textstyle\frac{g}{2}}\int |\varphi_k|^4(x)\,dx\,.
\end{align}
This is the celebrated Bose-Hubbard model \cite{Fish89b} restricted to three sites.
The coupling constant $g$ describes the two-body interactions. 
In order to avoid degeneracies in $\lf_k$, we assume a linear dependence on the additional
parameter $\oB$. In 
experimental realizations this can be achieved e.g.\ by a Stark field.
The zero point of the energy can always be shifted in order to fulfill $\bar\lf=0$. 
We fix the values of 
these constants to $\oB=0.1$, $\Kab_{12}=\Kab_{23}=0.25$ and $\Knn=0.1$ in order to 
describe the same 
situation as in \cite{06semiBEC}. These parameter values are experimentally accessible with standard 
techniques and avoid degeneracies in the Hamiltonian \eqref{eq-Hqm} so that the present study shows 
the most general diversification.


We consider a system of $N=30$ particles. The many particle Hilbert space has 
dimension $L = (N+1)(N+2)/2$ and the eigenenergies $E_k$ and eigenstates $\ket{\ef_k}$, $k=1,...,L$, can be calculated numerically exactly in the Fock basis $\ket{\vec{n}}$ ($\vec{n}=(n_1,n_2,n_3))$,
\begin{equation}\label{eq-eigexpansion}
\ket{\ef_k}\,= \!\!\!\!\!\!\sum_{n_1+n_2+n_3=N}\!\!\!\!\!\!c^{(k)}_{\vec{n}}\,\ket{\,\vec{n}\,}
\end{equation}
with $c^{(k)}_{\vec{n}}=c^{(k)}_{n_1,n_2,n_3}\in\R$.
In \cite{06semiBEC} we introduced a dynamical (semiclassical) representation $\ket{\,\vec{\varphi}\,}$ 
with $\vec\varphi=(\varphi_1,\varphi_2,\varphi_3)\in T^3$ in order to classify the eigenstates
$\ket{\ef_k}$ by the dynamical behavior of the corresponding classical system. We divided all
eigenstates $\ef_k(\vec\varphi)$ into six categories by assigning new geometrical quantum numbers to
the states with respect to their localization properties on the toroidal configuration space $T^3$. In the
dynamical representation \cite{Sibe96}, the eigenstates can be written as
\begin{equation}\label{eq-eigf}
\ef_k(\vec{\varphi}) = \scal{\vec\varphi\,}{\,\ef_k} = \!\!\!\!\!\!\!\sum_{n_1+n_2+n_3=N}\!\!\!\!\!\!\!
c^{(k)}_{\vec{n}}\,e^{i\vec n \cdot\vec\varphi}\,.
\end{equation}
Many eigenstates can be directly assigned to a category (A--E1) with the idealized functional forms
\noindent
\begin{eqnarray}
\label{eq-drepA}
\scal{\vec\varphi\,}{\l_1,\l_2}_A &\!\!=\!\!& 
	\eta^3\,e^{i\l_1\varphi_1}\,e^{i(N-\l_1-\l_2)\varphi_2}\,e^{i\l_2\varphi_3}\,,\\
\label{eq-drepB}
\scal{\vec\varphi\,}{\l,\t}_B &\!\!=\!\!& 
	\eta^2\, e^{i\l\varphi_1}
	\,e^{i(\alpha_1\varphi_2+\alpha_2\varphi_3)}\,\chi_\t(\varphi_3-\varphi_2)\,,\\
\label{eq-drepC}
\scal{\vec\varphi\,}{\l,\t}_C &\!\!=\!\!& 
	\eta^2\, e^{i(\alpha_1\varphi_1+\alpha_2\varphi_2)}\,e^{i\l\varphi_3}
	\chi_\t(\varphi_1-\varphi_2)\,,\\
\label{eq-drepD}
\scal{\vec\varphi\,}{\l,\t}_D &\!\!=\!\!& \eta^2\, 
	e^{i\l(\varphi_1+\varphi_3)/2}\,e^{i(N-\l)\varphi_2}\,\chi_\t(\varphi_1-\varphi_3),\\
\label{eq-drepE1}
\scal{\vec\varphi\,}{\t_d,\t_a}_{E1} &\!\!=\!\!&2^{-1/2}\eta
	 \,e^{iN(\varphi_1+\varphi_2+\varphi_3)/3}\nonumber\\[-1mm]&&\quad
	\chi_{\t_d}(\varphi_1+\varphi_3-2\varphi_2)\,
	\chi_{\t_a}(\varphi_1-\varphi_3)\,,\hspace{0.7cm}
\end{eqnarray}
where $\eta\!=\!(2\pi)^{-1/2}$ is a normalization constant, $\l$, $\l_1$, $\l_2$, $\t$, $\t_a$, $\t_d\in\N_0$ are quantum numbers  and $\chi_n$ is a harmonic oscillator 
eigenfunction, which is localized in the configuration space $[-\pi,\pi]$. The sixth type (E2) 
consists of irregular superposition patterns and represents the chaotic regions 
of the classical phase space. We will not consider these states here, but study only the properties of such 
eigenstates that are qualitatively similar to one of the functions of
 Eqs.~\eqref{eq-drepA}--\eqref{eq-drepE1}.
In a good approximation, the parameters $\alpha_i\in\N_0$ with $\sum\alpha_i=N-\l$ only depend on the quantum number $\l$. 
If an eigenstate separates in the variable $\varphi_k$, i.e.\ $\varphi_k$ enters only in the form $e^{in_k\varphi_k}$, then the 
condensate in well $k$ is decoupled from the rest and is an eigenstate of the 
number operator in semiclassical approximation ($\hbar=1$)
\begin{equation}\label{eq-nkdef}
\On_k=-i\diff{}{\varphi_k}\,.
\end{equation}
I.e.\ it has a definite 
number of particles. The representation \eqref{eq-nkdef} of the number operators 
$\On_k=\Oaa_k\Oa_k$ can be verified 
by comparing Eqs.~\eqref{eq-eigexpansion} and \eqref{eq-eigf}. So, $\On_k$ acts as a {\it momentum operator} in the dynamical representation.

In type (D), e.g., site $2$ is decoupled from the rest and has the fixed number of particles $N-\l$ while sites $1$ and $3$ are entangled with the total number of particles $\l$. 

The overall picture is described as follows: There cannot be assigned global quantum numbers to all eigenstates due to the lack of symmetries in the system. However, the whole Hilbert space can be divided into various sectors, where one of the sets of geometric quantum numbers of Eqs.~\eqref{eq-drepA}--\eqref{eq-drepE1} applies and characterizes the eigenstates completely.


%
\begin{figure}[htb]
\begin{center}
\vspace{-2mm}
\includegraphics[width=7.0cm]{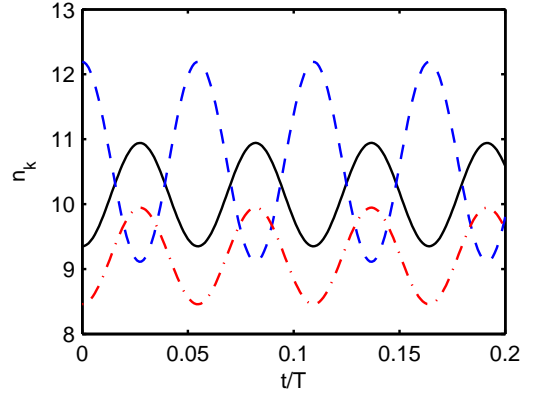}
\end{center}
\vspace{-1mm}
\caption{\label{fig-punkt_0_1}(Color online) Time evolution of the state $\ket{\Psi}=(\ket{0,0}_E
+ \ket{1,0}_E)/\sqrt{2}$. Shown are the particle numbers in well $1$ (black, solid), in well $2$ 
(blue, dashed), and in well $3$ (red,dash-dotted). The time is measured in multiples of 
$T=2\pi/\oB$.}
\end{figure}
Now, we show how easy one can manipulate and understand the dynamical behavior of 
Bose-Einstein condensates using the dynamical representation \eqref{eq-drepA}--\eqref{eq-drepE1}. 
The easiest way to get 
a nontrivial dynamics is to superpose two eigenstates $\ket{\ef_a}$ and $\ket{\ef_b}$ of the system 
and study the time evolution:
\begin{equation}
\ket{\Psi(t)} = c_a\,e^{-iE_at}\,\ket{\ef_a}+c_b\,e^{-iE_bt}\,\ket{\ef_b}
\end{equation}
with $c_{a/b} = |c_{a/b}|\,e^{-i\gamma_{a/b}}$ and $a,b\in\{1,...,L\}$.
The number operators $\On_k$ are fundamental observables for a Bose-Einstein condensate. Their time evolution is given by $\ex{n_k}_t = \scal{\Psi(t)}{\On_k|\Psi(t)}$, which can be calculated as
\begin{eqnarray}
\ex{\On_k}_t &=& |c_a|^2\scal{\ef_a}{\On_k|\ef_a} +|c_b|^2\scal{\ef_b}{\On_k|\ef_b} \nonumber
 \\&&+  2|c_a|\,|c_b|\,|\scal{\ef_a}{\On_k|\ef_b}|\,\cos(\Omega t+\Delta\gamma+\xi_k)\,.\hspace{0.7cm}
\end{eqnarray}
Here, $\Omega=E_a-E_b$, $\Delta\gamma=\gamma_a-\gamma_b$ and we used the decomposition 
$\scal{\ef_a}{\On_k|\ef_b}=|\scal{\ef_a}{\On_k|\ef_b}|\,e^{i\xi_k}$. The oscillation amplitude is 
proportional to the modulus of $\scal{\ef_a}{\On_k|\ef_b}$, which we can calculate directly using 
Eqs.~\eqref{eq-drepA}--\eqref{eq-drepE1}. For the type (A) eigenstates, all three wells are 
decoupled, i.e.\ are eigenstates of all three number operators $\On_k$. Therefore, the non-diagonal 
matrix elements are zero and the behavior is time independent. Since the behavior of type (B) is 
equivalent to that of type (C), we discuss only types (C), (D) and (E1). For convenience we choose $|c_{a/b}|=1/\sqrt{2}$.

Type (E1) shows entanglement between all three wells. The matrix elements 
$\scal{\t_a',\t_b'}{\On_k|\t_a,\t_b}_{E_1}$ can be calculated analytically using Eqs.~\eqref{eq-drepE1} and \eqref{eq-nkdef}. First, the integration region can be 
extended from the torus $T^3$ to the the whole $\R^3$ due to the localization of the oscillator
functions. Second, the matrix elements can be evaluated directly by transforming to new variables 
$x_d = \varphi_1-2\varphi_2+\varphi_3$, $x_a = \varphi_1-\varphi_3$, $x_3 = \varphi_2$,
and representing the new differential operators $-i\partial/\partial x_s$ by creation and annihilation operators,
\begin{equation} \label{eq-bdef}
-i\diff{}{x_s}\,\chi_{\t_s}(x_s) = -i\sqrt{\frac{\mf_s\Omega_s}{2}}\,(\hat b_s - \hat b^\dagger_s)\,\,\chi_{\t_s}(x_s)
\end{equation}
with $s = d,a$.
The frequencies $\Omega_{d/a}$ are given by 
the energy differences for the diagonal resp. antidiagonal degree of freedom of the local harmonic 
oscillators with energies $\epsilon_{d/a}^{(j)}=\Omega_{d/a}(j+1/2)$. 
The effective mass $\mf_{d/a}$ is the length scaling for the eigenfunctions
$\scal{\vec\varphi\,}{\t_d,\t_a}$.
\begin{figure}[htb]
\begin{center}
\includegraphics[width=7.0cm]{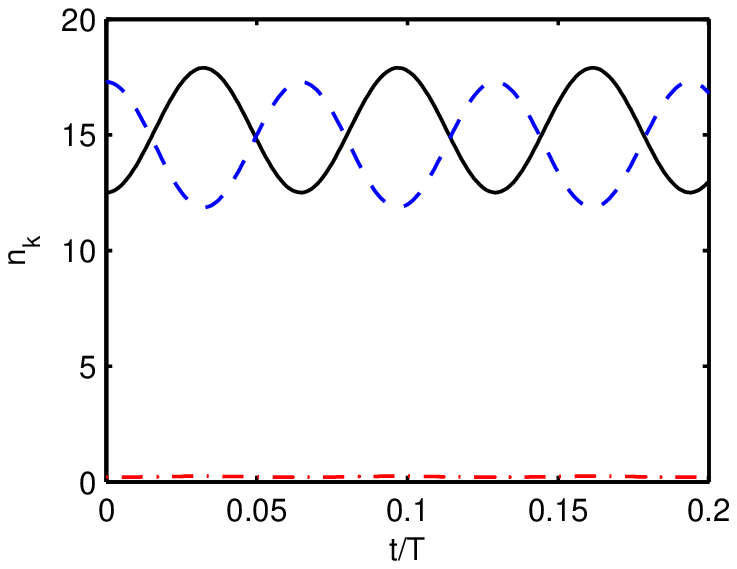}
\end{center}
\caption{\label{fig-Linie_1_0_4}(Color online) Time evolution of the state $\ket{\Psi}=(\ket{0,3}_C
+ \ket{0,4}_C)/\sqrt{2}$ and otherwise as in Fig.~\ref{fig-punkt_0_1}.}
\end{figure}
The matrix elements of $\On_k$ are different from zero only if one quantum number coincides 
while the other differs by $0,\pm1$. We consider the case that $\t_a=\t_a'$. 
Then, the diagonal matrix elements with 
$\t_d'=\t_d$ are given by $\scal{\t_d,\t_a}{\On_k|\t_d,\t_a}_{E1}= N/3$ for $k=1,2,3$ and the
non-diagonal elements yield for $\t_d'=\t_d\pm1$
\begin{eqnarray}
\scal{\t_d',\t_a}{\On_1|\t_d,\t_a}_{E1}&\!\!=\!\!&\phantom{-2}\,i\,\sgn(\t'\!-\!\t)\,
\sqrt{\frac{\mf_{d}\,\Omega_d\,\tm_d}{2}}\,,\\
\scal{\t_d',\t_a}{\On_2|\t_d,\t_a}_{E1}&\!\!=\!\!&
-2\,i\,\sgn(\t'\!-\!\t)\,\sqrt{\frac{\mf_{d}\,\Omega_d\,\tm_d}{2}}\,,\\
\scal{\t_d',\t_a}{\On_3|\t_d,\t_a}_{E1}&\!\!=\!\!& 
\phantom{-2}\,i\,\sgn(\t'\!-\!\t)\,
\sqrt{\frac{\mf_{d}\,\Omega_d\,\tm_d}{2}}\,.\hspace{0.7cm}
\end{eqnarray}
Here,  $\tm=\max(\t,\t')$ and $\sgn(\t)$ is the sign function with $\sgn(0)=0$.
The particle numbers in wells 1 and 3 oscillate in phase with the same amplitude, while $n_2$ oscillates with opposite phase.
Fig.~\ref{fig-punkt_0_1} shows an example for excitation numbers $\t_d=0,1$ and $\t_a=0$. Shown is the 
numerically exact dynamics of the particle numbers $n_k$ defined by
\begin{equation}
n_k(t) = \int_{\text{well}\;k}\ex{\OP^\dagger(x)\OP(x)}_t\,dx\approx\ex{\Oa_k^\dagger\Oa_k}_t\,,
\end{equation}
where the expectation value is calculated for the state $\ket{\Psi(t)}$. The dynamical representation of the eigenstates with the idealized wave 
functions \eqref{eq-drepA}--\eqref{eq-drepE1} describes the dynamical behavior not only
qualitatively but also 
quantitatively. With $m_{d}^\text{eff}=0.7$ and $\Omega_d=1.8$, the oscillation amplitude yields 
$\Delta n_{1/3}^\text{eff}=0.8$, which coincides with the exact value obtained from 
Fig.~\ref{fig-punkt_0_1}. The different offsets of the oscillations in Fig.~\ref{fig-punkt_0_1} are 
due to deviations of the phases of the exact eigenstates from the plane waves of the idealized wave 
functions~\eqref{eq-drepE1}.

\begin{figure}[h]
\begin{center}
\vspace{-2mm}
\includegraphics[width=7.0cm]{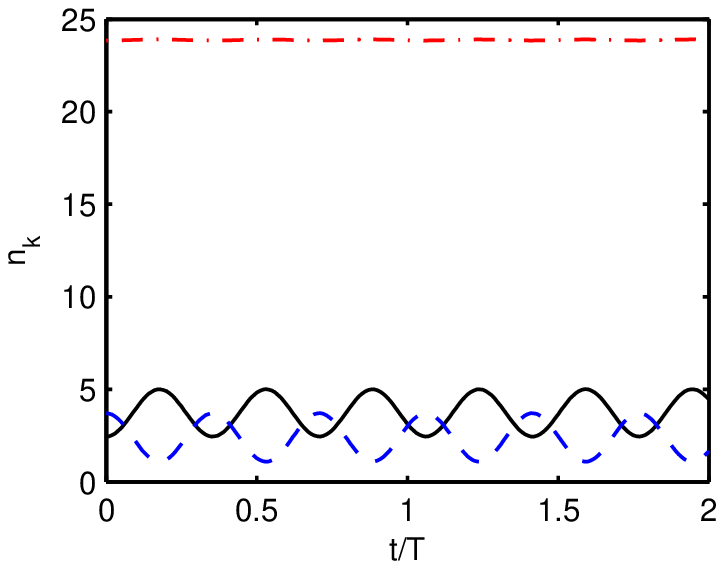}
\end{center}
\caption{\label{fig-Linie_1_24_5}(Color online) Time evolution of the state $\ket{\Psi}=(\ket{24,4}_C
+ \ket{24,5}_C)/\sqrt{2}$ and otherwise as in Fig.~\ref{fig-punkt_0_1}.}
\end{figure}

Beatings between eigenstates of type (C) are shown in Figs.~\ref{fig-Linie_1_0_4} and 
\ref{fig-Linie_1_24_5} for states $(\ket{0,3}_C+\ket{0,4}_C)/\sqrt{2}$ and
 $(\ket{24,4}_C+\ket{24,5}_C)/\sqrt{2}$ respectively. 
The matrix elements are calculated along the same idea as in the previous example:
First, one transforms into new coordinates $x_1 = \varphi_1-\varphi_2$, 
$x_2 = \varphi_3-\varphi_2$ and $x_3 = \varphi_2$, which maps the torus $T^3$ onto itself (shear). 
Then one extends the integration region in $x_1$-direction to the whole $\R$ and represents $-i\partial/\partial x_1$ as in Eq.~\eqref{eq-bdef}.
The matrix elements 
$\scal{\l',\t'}{n_{k}|\l,\t}_C$ vanish for $\l'\neq\l$ and also for $|\t-\t'|>1$. So, the 
non-zero elements are ($r=1,2$, $\,\t'-\t=0,\pm1$)
\begin{eqnarray}
\scal{\l,\t'}{n_{r}|\l,\t}_C&\!\!\!=\!\!\!&\alpha_{r} \,\delta_{\t'\!,\t}\!+\!(\!-1)^{r}\sgn(\t\!-\!\t')\sqrt{\frac{\mf\,\Omega\,
 \tm}{2}}\,,\hspace{0.6cm}\\
\scal{\l,\t'}{n_3|\l,\t}_C &\!\!\!=\!\!\!& \l\,\delta_{\t'\!,\t}\,.
\end{eqnarray}
\\
Fig.~\ref{fig-Linie_1_0_4} shows particle oscillations between wells $1$ and $2$, while the number of particles in well $3$ is constant with $n_3=0.2$. This is the effect of self-trapping directly shown in the quantum regime. The example shows the typical behavior of beatings in this class of eigenstates: A number of $n_3=\l$ particles is trapped in well $3$, while the rest is oscillating coherently between the other wells with an amplitude proportional to the square root of the transversal quantum number. With $\mf=2.1$ and $\Omega=1.8$, the oscillation amplitude yields a value of $\Delta n^\text{eff}_{1/2}= 2.7$, which coincides with the exact value taken from 
Fig.~\ref{fig-Linie_1_0_4}.
The second example of this type in Fig.~\ref{fig-Linie_1_24_5} shows the opposite regime where 
nearly all particles are localized in well $3$ and the rest oscillates with amplitude 
$\Delta n_{1/2}=1.3$. The matrix element yields for the amplitude the value 
$\Delta n^\text{eff}_{1/2}=1.2$ ($\mf=1.8$, $\Omega=0.3$).

\begin{figure}[htb]
\begin{center}
\includegraphics[width=7.0cm]{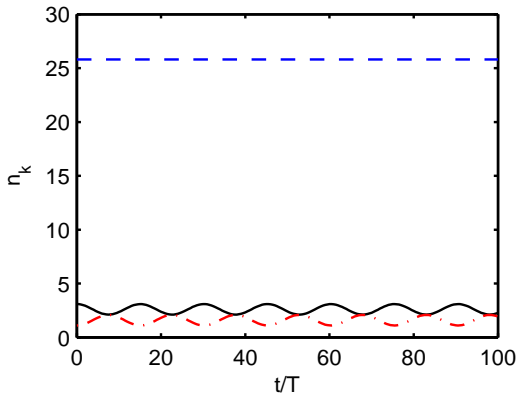}
\end{center}
\caption{\label{fig-Linie_12_4_1}(Color online) Time evolution of the state $\ket{\Psi}=(\ket{4,0}_D
+ \ket{4,1}_D)/\sqrt{2}$ and otherwise as in Fig.~\ref{fig-punkt_0_1}.}
\end{figure}

\begin{figure}[htb]
\vspace{-3mm}
\begin{center}
\includegraphics[width=7.0cm]{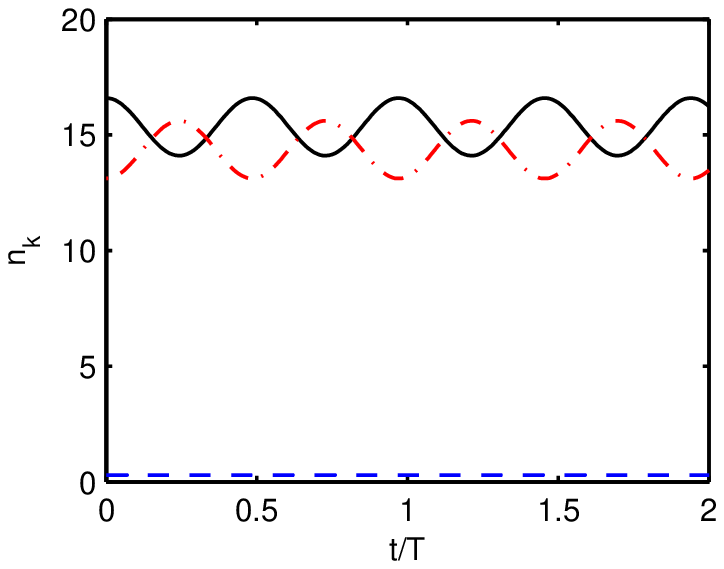}
\end{center}
\caption{\label{fig-Linie_12_30_3}(Color online) Time evolution of the state $\ket{\Psi}=(\ket{30,2}_D
+ \ket{30,3}_D)/\sqrt{2}$ and otherwise as in Fig.~\ref{fig-punkt_0_1}.}
\end{figure}

In type (D), wells $1$ and $3$ are phase locked, although there does not exist a direct coupling term in the Hamiltonian~\eqref{eq-Hqm}. Both wells couple indirectly through the middle well, however also here we can completely decouple the dynamics. The particle dynamics of the state 
$(\ket{4,0}_D+ \ket{4,1}_D)/\sqrt{2}$ is shown as an example in Fig.~\ref{fig-Linie_12_4_1}. In order to understand the behavior, we again have to evaluate the matrix elements 
$\scal{\l',\t'}{\On_k|\l,\t}$. The non-vanishing elements with $\l'=\l$ and $|\t'-\t|\leq1$ can be written as ($r = 1,3$, $\,\t'-\t=0,\pm1$)
\begin{eqnarray}
\scal{\l,\t'}{n_{r}|\l,\t}_D&\!\!\!=\!\!\!&{\textstyle\frac{\l}{2}}\,\delta_{\t'\!,\t}
\!+\!i(2\!-\!r)\sgn(\t'\!-\!\t)
\sqrt{\frac{\mf\Omega \tm}{2}}\,,\hspace{0.5cm}\\
\scal{\l,\t'}{n_2|\l,\t}_D &\!\!\!=\!\!\!& (N-\l)\,\delta_{\t'\!,\t}\,.
\end{eqnarray}
Interesting are the two extremes, where all the population is in or outside of the middle well. In 
Fig.~\ref{fig-Linie_12_4_1}  nearly  all particles ($n_2 = 25.8$) are concentrated in well 2 and 
the population of the others oscillates between them through the middle well. Calculation of the 
oscillation amplitudes by the matrix elements yields a value of 
$\Delta n_{1/3}^\text{eff}=0.46$ ($\mf=0.7$, $\Omega=0.6$) in 
comparison with the exact value of $\Delta n_{1/3}=0.49$. I.e., one particle is oscillating between the outer 
wells on average.
The most impressive example is shown in Fig.~\ref{fig-Linie_12_30_3}. The occupation of the middle 
well with $n_2 = 0.2$ is much smaller than the  oscillation amplitude between the first and third 
well with $\Delta n_{1/3}=1.3$. Although the middle well is effectively not occupied, there 
is a particle oscillation of more than two particles through this well. Also here, 
the matrix element of the idealized wave functions yields a very accurate value 
for the oscillation amplitude of $\Delta n^\text{eff}_{1/3} = 1.4$ ($\mf=2.8$, $\Omega=0.5$).\\
%

Concluding, we have shown various types of oscillation dynamics by superposing eigenstates.
The classification of eigenstates in the dynamical representation provides a powerful tool in order to understand and manipulate the behavior of condensates in a few-well potential.
For a larger number of particles $N$, there exist longer ladders of transversal excitations in each class leading to higher transversal quantum numbers. Accordingly, the amplitude of the particle oscillations can be increased by superimposing such higher excited states.
There is also a generalization to longer chains of wells with oscillations between more distant wells, more on this in future work.


\end{document}